\documentclass[twocolumn,showpacs]{revtex4-1}
\usepackage{graphicx}
\usepackage{amsmath}
\usepackage{amsfonts}
\usepackage{amssymb}
\usepackage{enumerate}
\usepackage{subfig}
\usepackage{epsfig}
\usepackage{dcolumn}
\usepackage{bm}

\newcommand{\be}{\begin{equation}}
\newcommand{\ee}{\end{equation}}
\newcommand{\bn}{\begin{eqnarray}}
\newcommand{\en}{\end{eqnarray}}

\begin{document}

\title{Phase Transition in IrTe$_{2}$ induced by spin-orbit coupling}

\author{S. Koley$^{1,2}$}
\affiliation{
$^{1}$Department of Physics and Meteorology,
Indian Institute of Technology, Kharagpur, 721302, India\\
$^{2}$ Present Address:Department of Physics, St. Anthony's College, Shillong, 
Meghalaya, 793001, India
}

\begin{abstract}
\noindent IrTe$_2$ has been renewed as an interesting system showing competing phenomenon between a questionable density-wave transition near 270 K followed by 
superconductivity with doping of high atomic number materials. Higher  
atomic numbers of Te and Ir supports strong spin-orbital coupling in this 
system. Using dynamical mean field theory with LDA band structure 
I have 
introduced Rashba spin orbit coupling in this system to get the interpretation 
for anomalous resistivity and related transition in this system. While no 
considerable changes are observed in DMFT results of Ir-5d band other than 
orbital selective pseudogap `pinned' to Fermi level, Te-p band shows a van Hove 
singularity at the Fermi level except low temperature. Finally I discuss the 
implications of these results in theoretical understanding of ordering in IrTe$_2.$  
\end{abstract}

\maketitle

\section{Introduction}
\noindent Transition-metal compounds are found to have interesting 
physical properties such as spin-charge-orbital density waves,\cite{at1,tokura} 
superconductivity\cite{sipos,sk,shen} at moderate to low temperature, unconventional 
transport and magnetism\cite{motohashi}. 
Most of these properties are believed to be originated from the topology of 
multiband Fermi surfaces\cite{yokoya} of the parent compound. The multi-orbital 
transition-metal oxides and chalcogenides 
exhibit metal-insulator transitions in presence of density wave orders induced 
instability in the Fermi surface \cite{imada}.        
Also five 3d orbitals of Fe play an important role in                       
 magnetism and superconductivity (SC) of Fe pnictides\cite{kontani,Johnston} and chalcogenides
like SC ordering in LaFeAsO$_{1-x}$F$_x$ \cite{Li,Mazin}. 
Along with this ``quantum fluctuation" is another 
parameter which introduce novel electronic states near zero temperature \cite{at3}. 
Failure of the Fermi liquid theory at magnetic quantum critical 
points (QCPs) in itinerant magnets\cite{vojta}, transition-metal 
oxides\cite{varma}, chalcogenides and heavy fermion systems\cite{coleman} are examples of the 
extant variety of quantum 
fluctuations. Transition-metal compounds with geometrically frustrated lattice 
structure, such as triangular lattices, which often 
form a ordered low temperature valence-bond solid (VBS)\cite{read} indeed show 
quantum fluctuation \cite{pwa}.

\noindent Recently, interplay between superconductivity and distortion in 
lattice structure in a transition metal dichalcogenide IrTe$_2$ of 1T structure 
have been noticed \cite{Pyon,kamitani}. 
It exhibits a phase transition at 270 K from the trigonal (P3m-1) 
to the monoclinic (C2/m) structure associated with related changes in 
electrical resistivity and magnetic susceptibility \cite{Fang}. 
In addition superconductivity can also be found   
in this compound substituting Ir of the parent 
compound by Pt\cite{ootsuki} or Pd \cite{zhou}. Both IrTe$_2$ and PtTe$_2$ 
have trigonal CdI$_2$-type structure and belong to the 
space group P3m-1 but Pt is having higher atomic number than Ir. High atomic 
number can be related to strong spin orbit (SO) coupling because it is directly 
proportional to the fourth power of atomic number. So the superconducting order in 
doped IrTe$_2$ is likely to be related to the spin orbit coupling. 
While undoped IrTe$_2$ manifests a first order  
phase transition whereas 
PtTe$_2$ shows no anomaly down to lowest temperatures.

\noindent In 1T-IrTe$_2$ the transition metal Ir layers and chalcogen Te layer
are bound to each other by interlayer van der Waals attraction with 
each of these layers forming a triangular lattice\cite{Yang}. So the transition metal 
ions are having three equivalent transition metal bonds.
The phase transition in IrTe$_2$ can be identified by an abrupt change in 
electrical resistivity and as well as a sudden decrease in Pauli paramagnetic 
susceptibility. 
Charge-orbital density order is predicted in Irte$_2$ at moderate to high 
temperature which results in strutural
 transition \cite{Yang}. Though a density wave transition is envisaged\cite{Yang},  
experiments manifest that one side of the triangular Ir lattice is changed 
uniformly whereas the other two Ir-Ir bonds remains same after the transition 
\cite{Pyon}. Moreover the parent chacogenides (IrTe$_2$ and PtTe$_2$) shows no 
sign of superconductivity down to lowest experimentally measured temperature\cite{raub,matthias}.
Due to large atomic number of Ir, the transition in the 
compound IrTe$_2$ is related with large SO coupling.  
Large SO coupling sometimes cause magnetic anisotropy and
multiferroicity\cite{Larson}. Furthermore it can lead to unconventional 
superconductivity\cite{Qi} in topological superconductors. Complete 
understanding of the electronic properties of IrTe$_2$ needs the 
explanation of structural phase transition.

\noindent Analysis of the extant experimental review speculates that IrTe$_2$
 with large SO coupling unveils almost bad metallic feature throughout 
all the temperature\cite{Fang}. The most intriguing phenomena is its phase 
transition near 270 K with anomaly in resistivity and magnetic 
susceptibility\cite{Fang}. The nature of this phenomena is still under debate. 
The existence of superconductivity with doping of Pd or Pt in association 
with disappearance of 270 K phase transition is another interesting 
phenomena in the phase diagram\cite{Pyon}. The phase diagram of IrTe$_2$ shows a huge 
resemblance with high T$_c$ cuprates and some heavy fermion superconductors\cite{haule}. 
The magnetic quantum critical point (QCP) is close to the superconductivity 
which gives the evidence of existance of a QCP of charge-orbital density wave 
nearer to the superconducting order and this density wave ordering may give 
rise to the superconducting order\cite{lonzarich}. Parent IrTe$_2$ shows multiband 
nature in its effective Fermi surface. Further modification of only one 
Ir-Ir bond\cite{Pyon} at the ordering temperature opposes the charge-orbital 
density wave order. Briefly all properties of IrTe$_2$ show strong correlation 
throughout all the temperature and doping, so dynamical mean field theory (DMFT)
 \cite{anisimov, kotliar} will be the best way to explain the unconventional 
ordering in it. In this paper I present a full DMFT calculation of IrTe$_2$ to 
predict the electronic properties of the system closer to the transition 
temperature region. 

\begin{figure}
\centering
(a)
\includegraphics[angle=270,width=0.8\columnwidth]{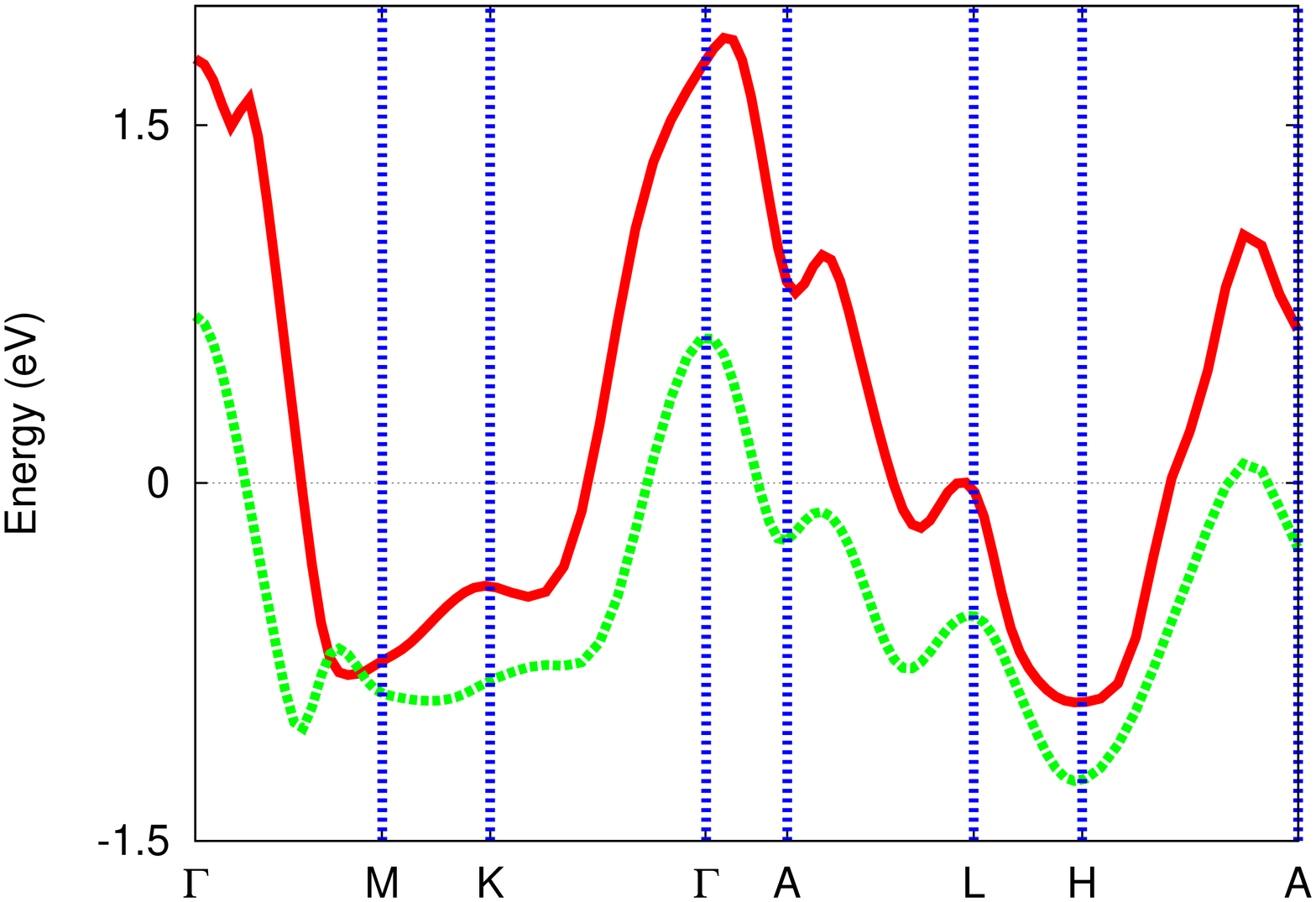}

(b)
\includegraphics[angle=270,width=0.8\columnwidth]{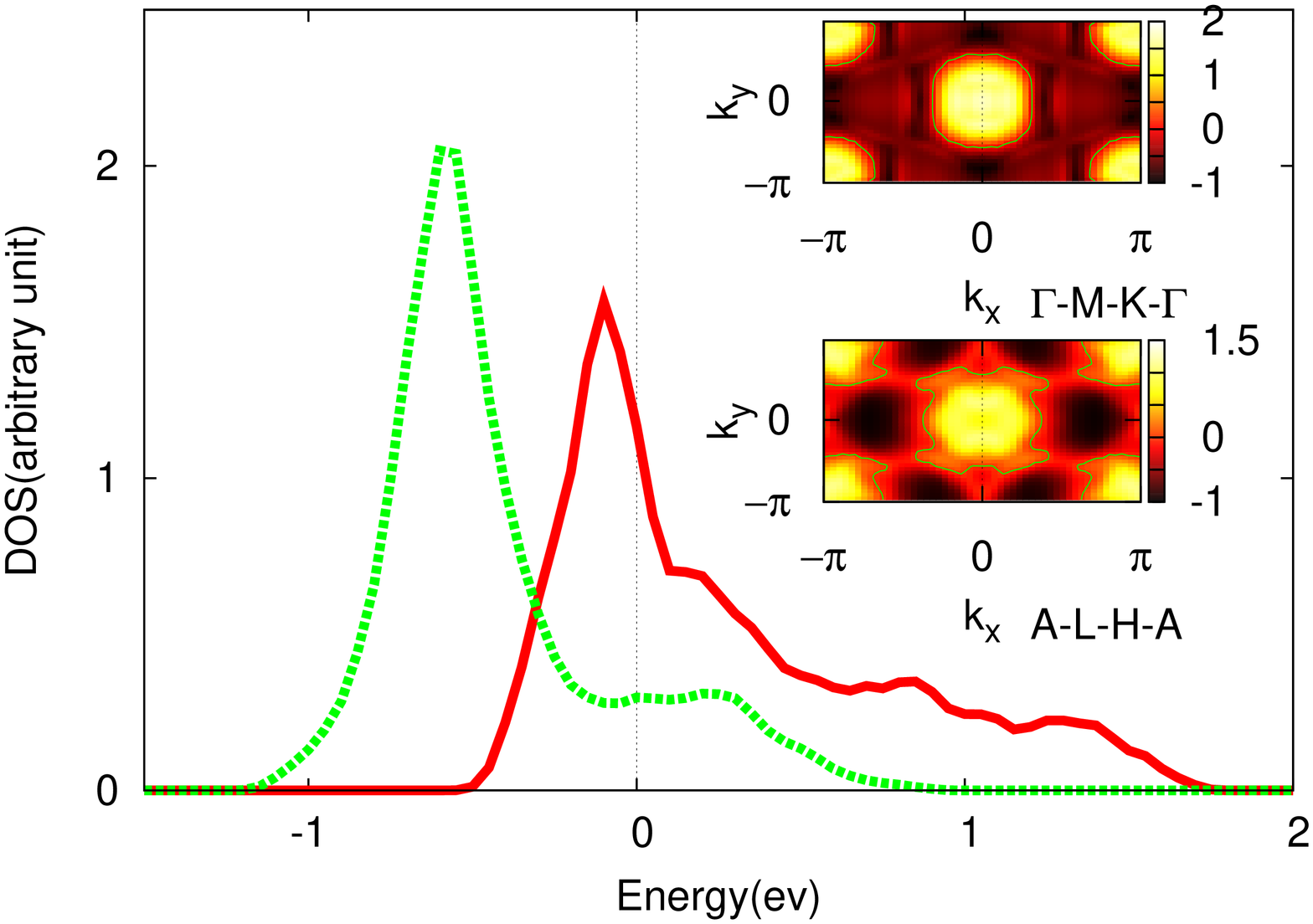}
\caption{(Color Online) (a)Ir-5d (red line) and Te-5p (Green line) bandstructure and (b) density of states near Fermi level. Fermi surface of IrTe$_2$
at two plane (inset of Fig.\ref{fig1}b). Upper inset is for the $\Gamma$-M-K-$\Gamma$ 
plane and lower inset is for A-L-H-A plane.}
\label{fig1}
\end{figure}

\noindent Tight binding band structure of IrTe$_2$ is calculated by using Ir-5d and 
Te-5p bands which gives two bands crossing Fermi energy level. Primarily these 
are the hybridized bands of Ir-5d and Te-5p orbitals. 
Although there is mixing of d and p orbital, Te-5p orbitals are playing 
dominant character in the bands near Fermi energy. LDA band structure 
calculation also predict the same. In Fig.\ref{fig1} I show the noninteracting
band structure, corresponding density of states and Fermi surface map (in 
$\Gamma -M -K -\Gamma$ plane and $A-L-H-A$ plane) of the 
bands nearest to the 
Fermi level. Band structure of IrTe$_2$ (Fig.\ref{fig1}a) clearly manifests the 
existence of negative 
indirect band gap. Six Fermi surface (FS) pockets can be seen from the LDA FS 
in Fig.\ref{fig1}b. The lower band is crossing Fermi level along $\Gamma$ - M direction 
closer to $\Gamma$ point, originated mainly from the Te-5p$_x$ 
and Te-5p$_y$ orbitals and is almost fullly filled while in the other band 
dominant contribution comes from Te-5p$_z$ and Ir-5d bands. The upper band 
is more spreaded and crosses Fermi level at higher momentum. Our band 
structure and FS calculation closely tracks the earlier results \cite{Yang}.

\noindent For IrTe$_2$, a two-band Hubbard model is mandated by LCAO results, 
and adequate treatment of dynamical correlations underlying bad metallic 
behavior is achieved by dynamical-mean field theory. DMFT approaches 
are proven to be successfull in treating strong dynamical fluctuations in 
correlated electronic systems \cite{at1,at2,at3,at4,at5} which made them useful 
tool of choice in the context of correlated electrons. 
The two band Hubbard model I used is (denoting lower band (Te-p) as `a' and 
upper band (Ir-d) as `b')
$H_{el}=\sum_{k,l,m,\sigma}(t_k^{lm}+\epsilon_l\delta_{lm})c_{kl\sigma}^{\dagger}c_{km\sigma}$ $+U\sum_{i,l=a,b}n_{il\uparrow}n_{il\downarrow}$ $+U_{lm}
\sum_in_{il}n_{im}.$ 
where l and m run over both band indices a and b. The intra-orbital 
correlation is taken as U 
and U$_{ab}$ is the inter-orbital correlation which play a major role 
throughout. Due to large atomic number of Ir inter-orbital hopping is 
negligible in comparision to spin orbit interaction. 
Moreover due to large spin-orbital coupling Rashba effect is 
an important addition to the two band Hubbard Hamiltonian. The Rashba 
Hamiltonian is $H_R = \alpha\sum_{k, \sigma, \sigma ^{'}}(\sigma \times p). \hat{z}c_{kl\sigma}^{\dagger}c_{kl\sigma ^{'}}$.
where $\alpha$ is coefficient of Rashba coupling, p is momentum and 
$\sigma$ is the Pauli matrix vector. To solve full Hamiltonian H=H$_{el}$+
H$_{Rashba}$ within 
DMFT I have combined the multi-orbital iterated perturbation theory 
(MOIPT) for H$_{el}$ 
\cite{laad} with a momentum-dependent modification to include Rashba 
coupling. The ordered state and physical observables are calculated
 self consistently from DMFT in presence of an infinitesimal symmetry breaking 
due to SO coupling.  
Following earlier procedure \cite{at1}, I will compute DMFT spectral functions and 
transport properties in the normal state and ordered state.

\begin{figure}
\centering
\includegraphics[angle=270,width=0.8\columnwidth]{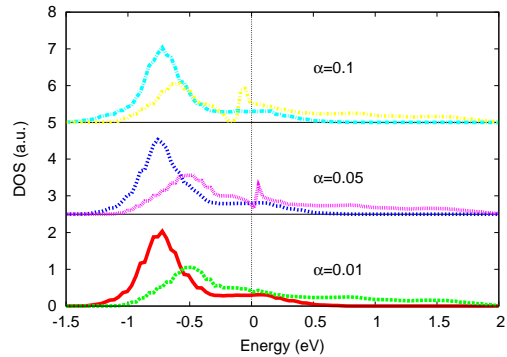}
\caption{(Color Online) Density of states near transition temperature 
(transition temperature is noted here for the temperature where resistivity 
shows peak value and around which hysteresis is observed) for 
different value of the coefficient of spin orbit coupling.}
\label{fig2}
\end{figure} 

\begin{figure}
\centering
(a)
\includegraphics[angle=270,width=0.8\columnwidth]{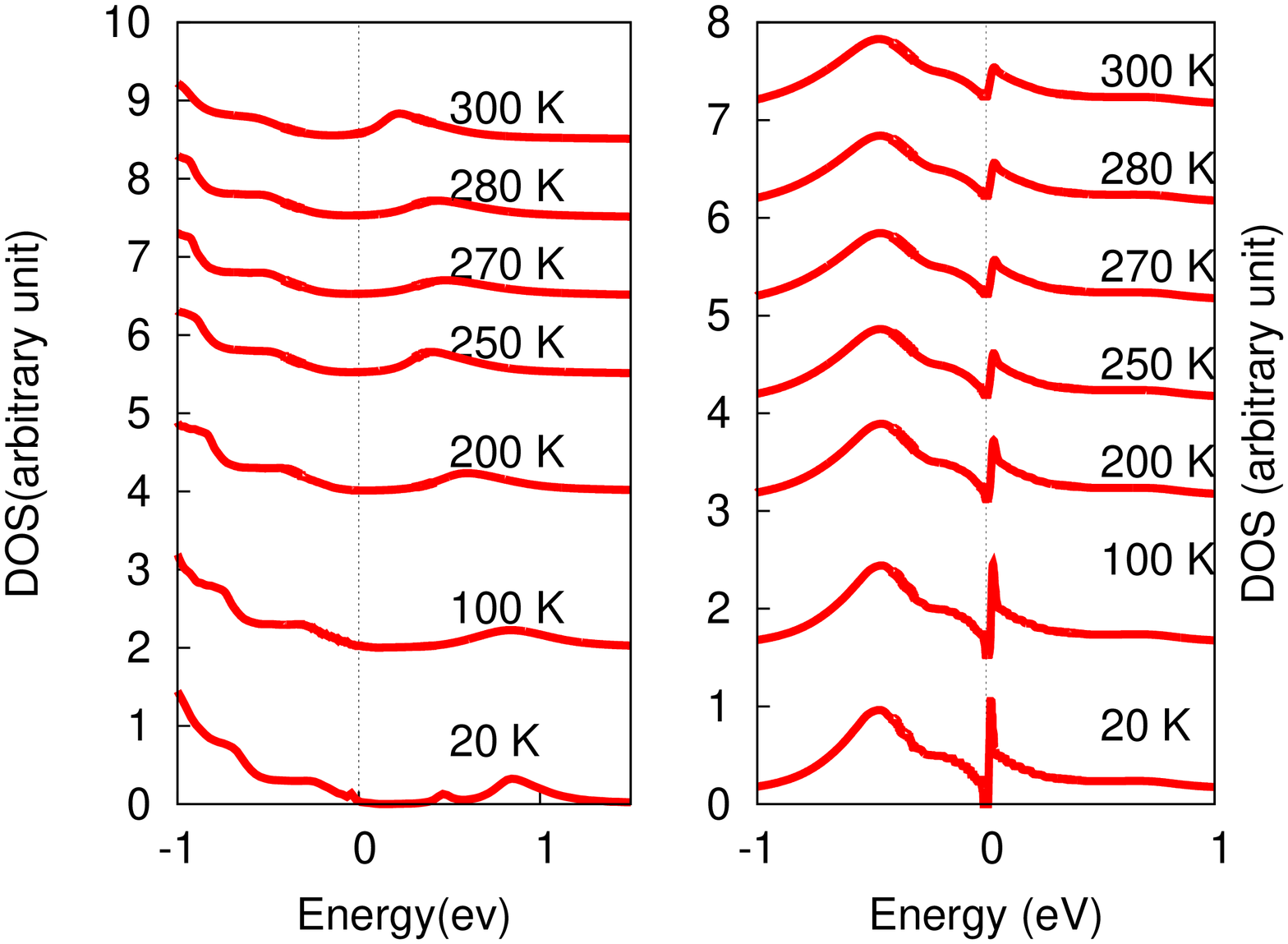}

(b)
\includegraphics[angle=270,width=0.8\columnwidth]{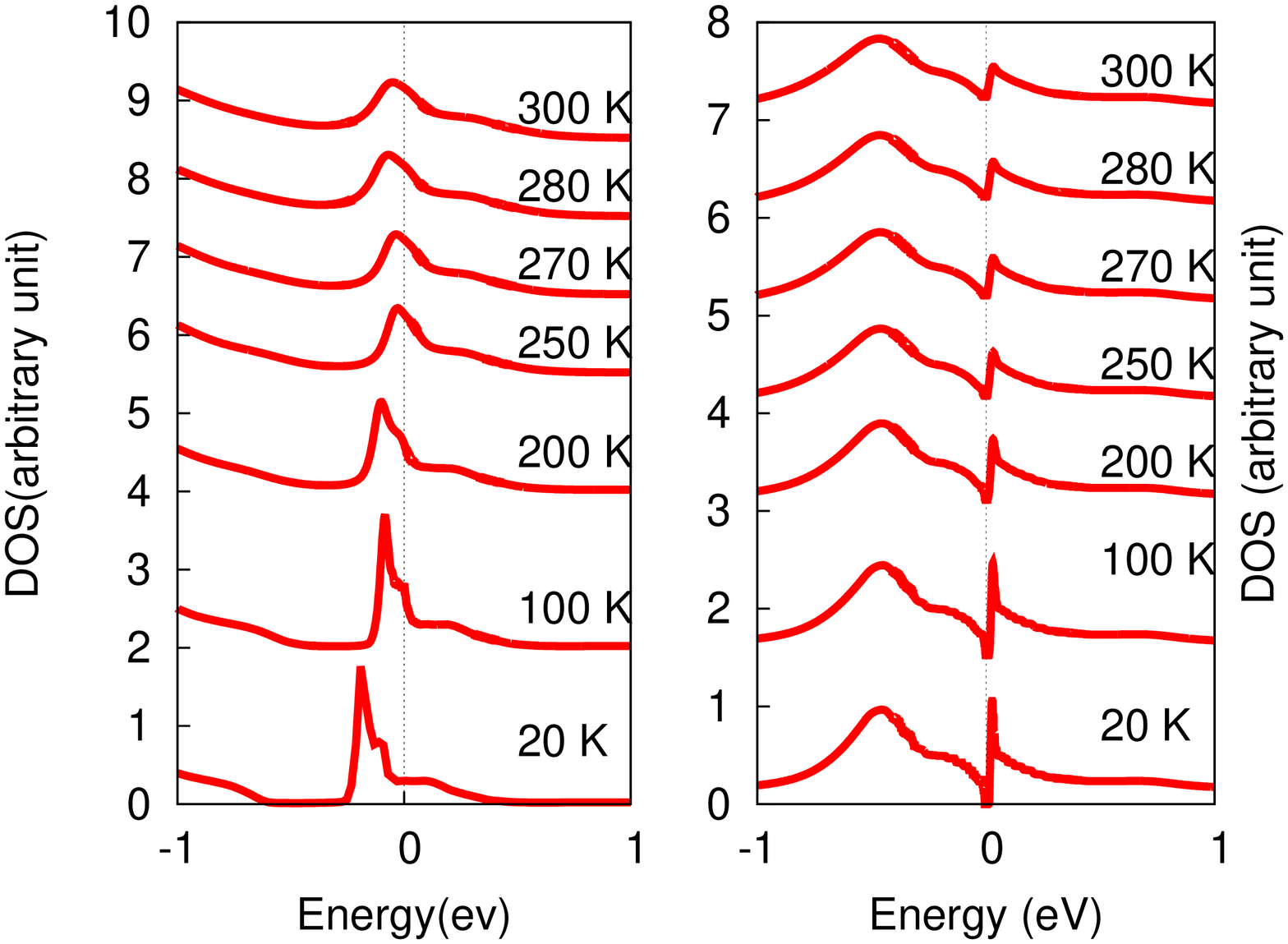}
\caption{(Color Online) $T$-dependent density of states for (a) temperature 
increasing from 0 K to 300 K, (b) temperature decreasing from 300 K to 0 K. 
Left panel is for Te-5p band and right panel is for Ir-5d band.}
\label{fig3}
\end{figure}
\begin{figure}
\centering
(a)
\includegraphics[angle=270,width=0.8\columnwidth]{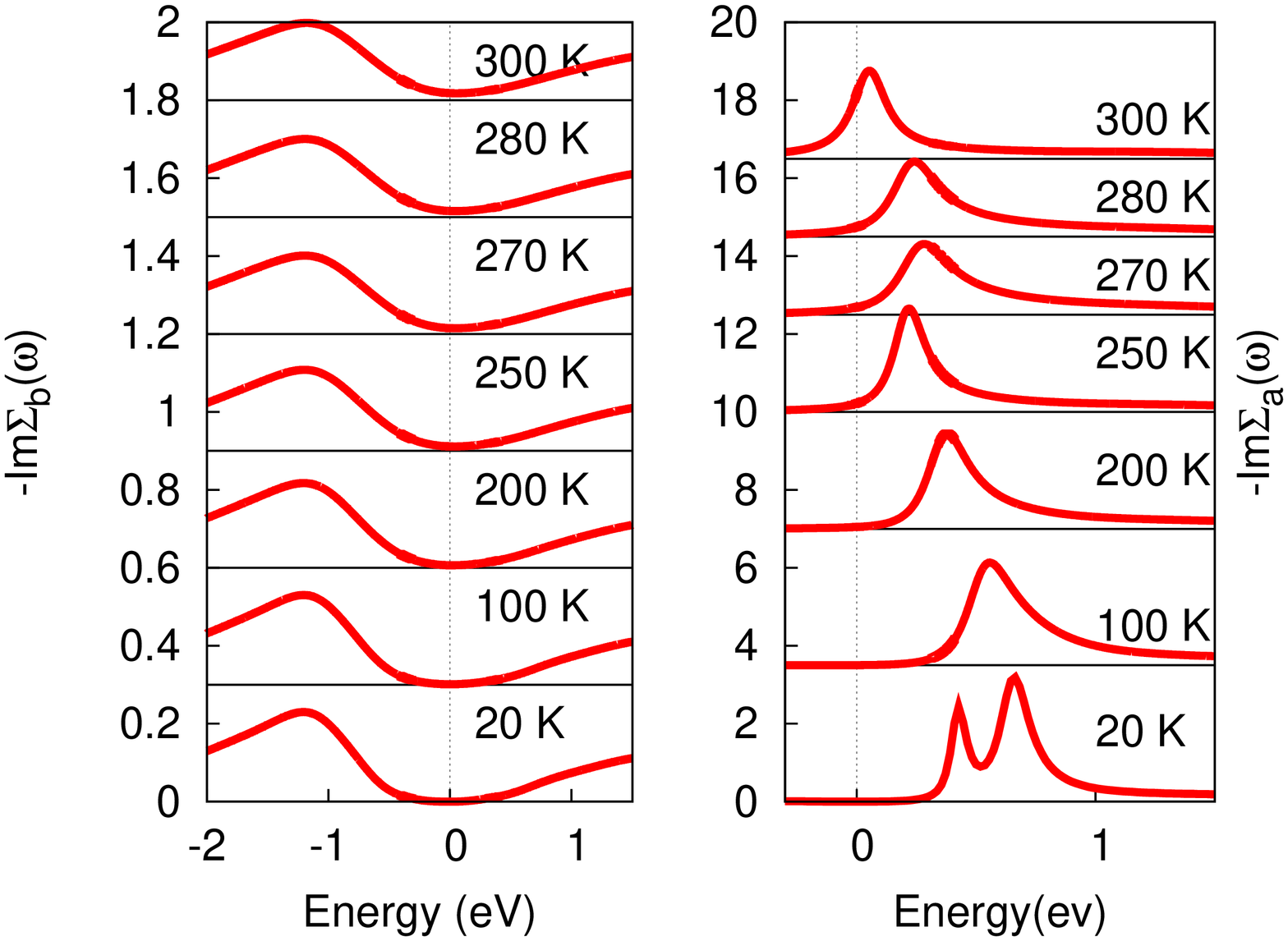}

(b)
\includegraphics[angle=270,width=0.8\columnwidth]{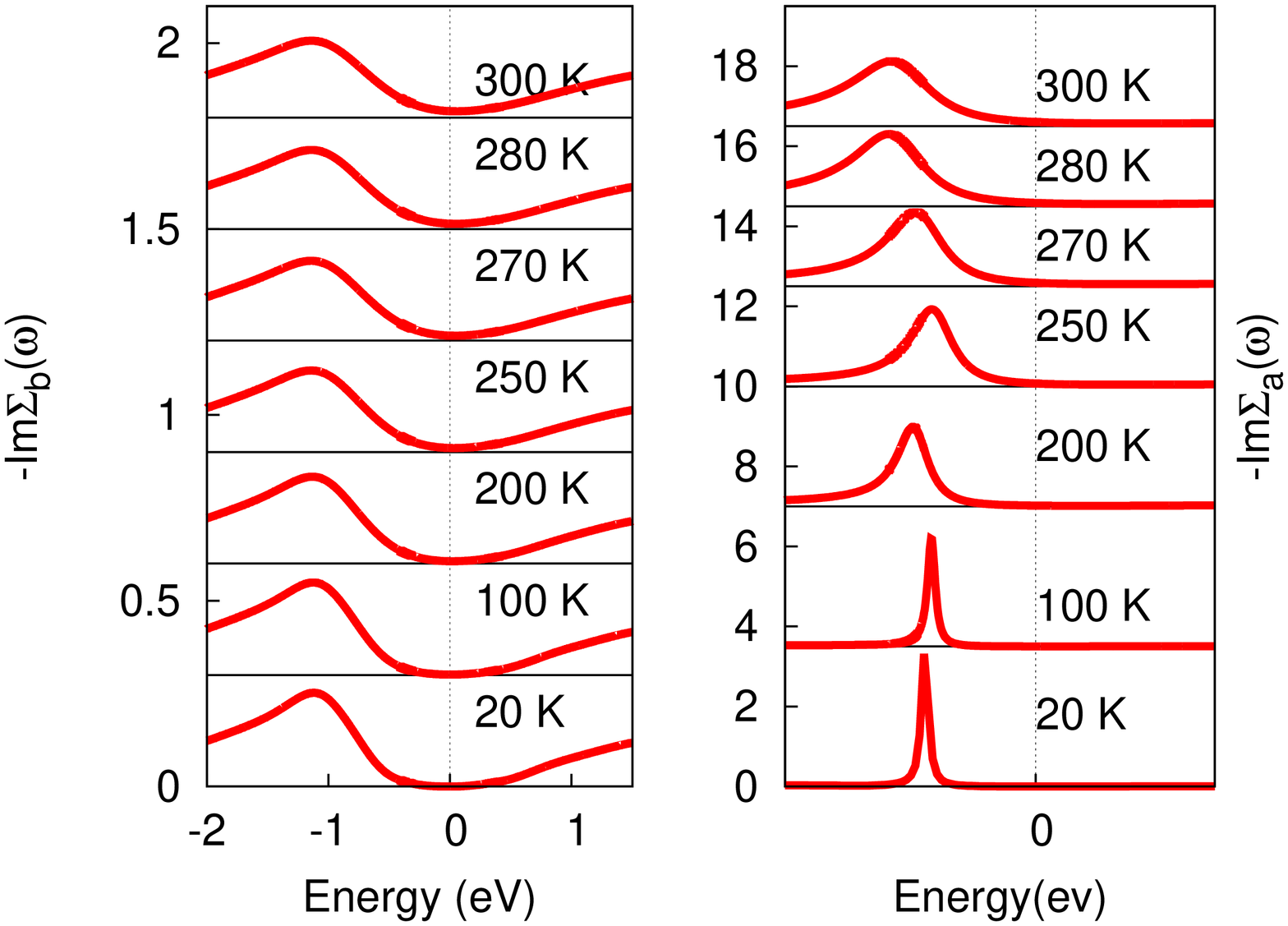}

(c)
\includegraphics[angle=270,width=0.8\columnwidth]{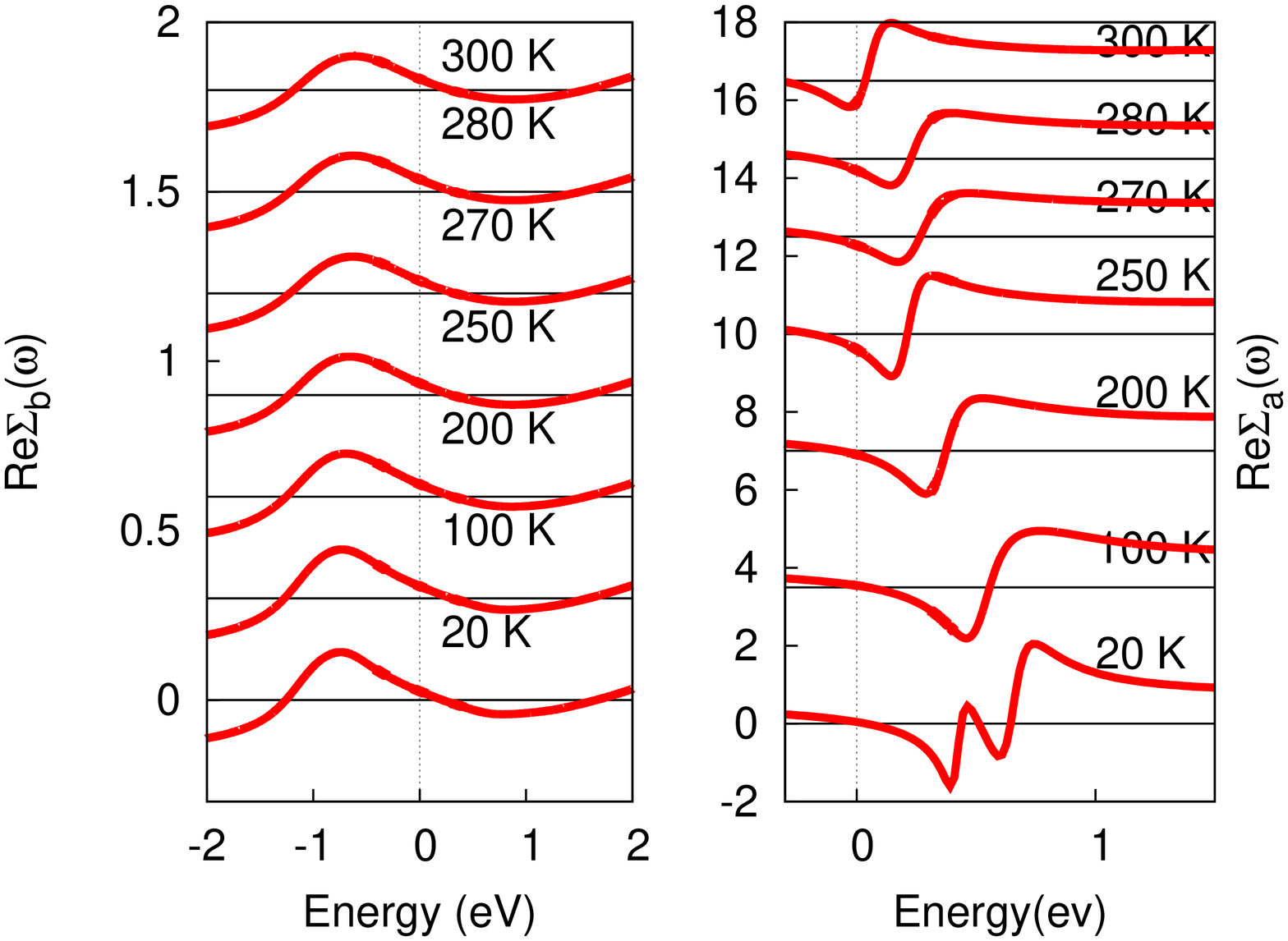}

(d)
\includegraphics[angle=270,width=0.8\columnwidth]{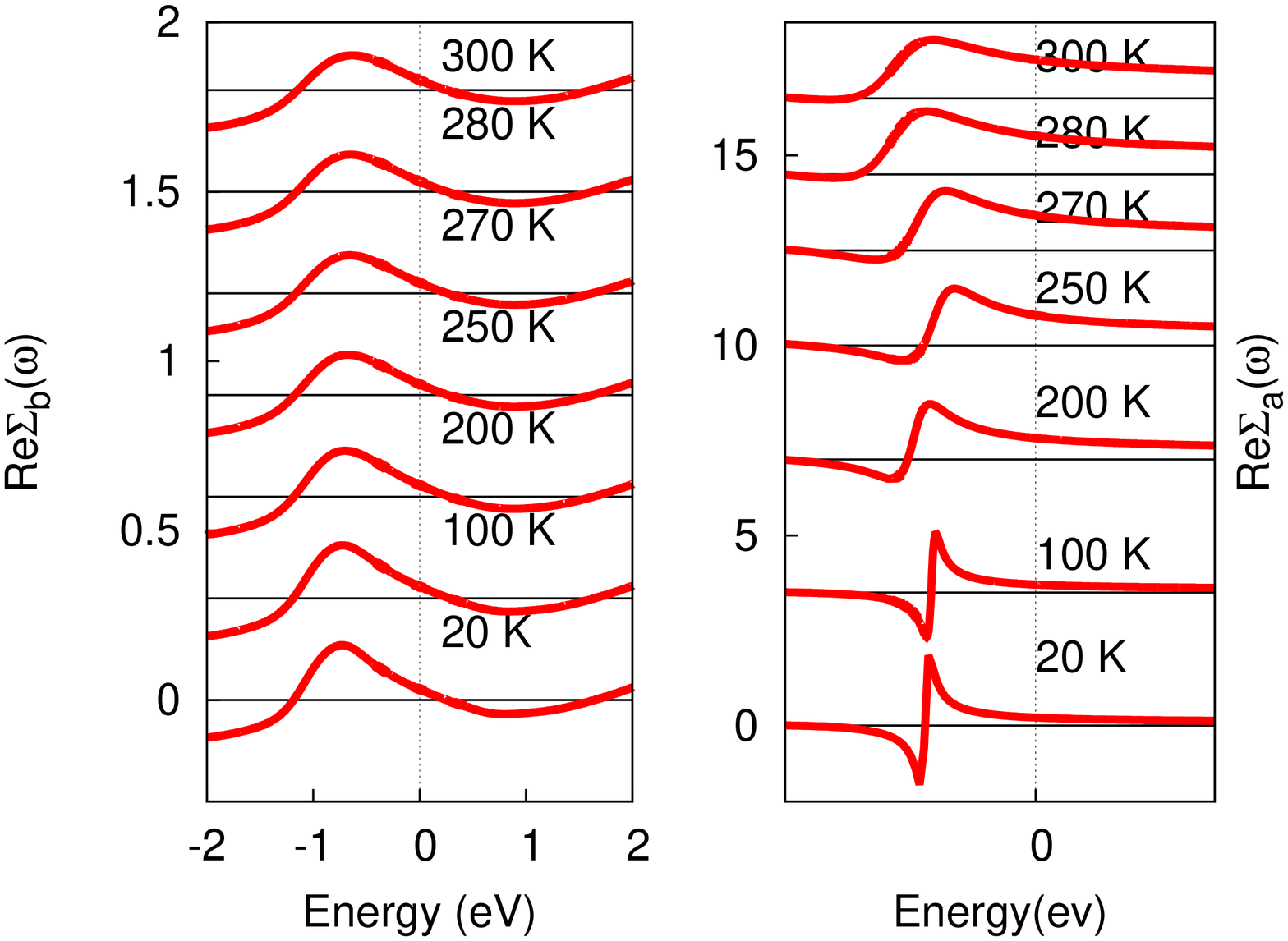}
\caption{(Color Online) $T$-dependent imaginary part of self energy when (a) 
temperature increasing from 0 K to 300 K, (b) temperature decreasing from 300 
K to 0 K. $T$-dependent real part of self energy when (c) temperature 
increasing from 0 K to 300 K, (d) temperature decreasing from 300 K to 0 K.}
\label{fig4}
\end{figure}
\begin{figure}
\centering
(a)
\includegraphics[angle=270,width=0.8\columnwidth]{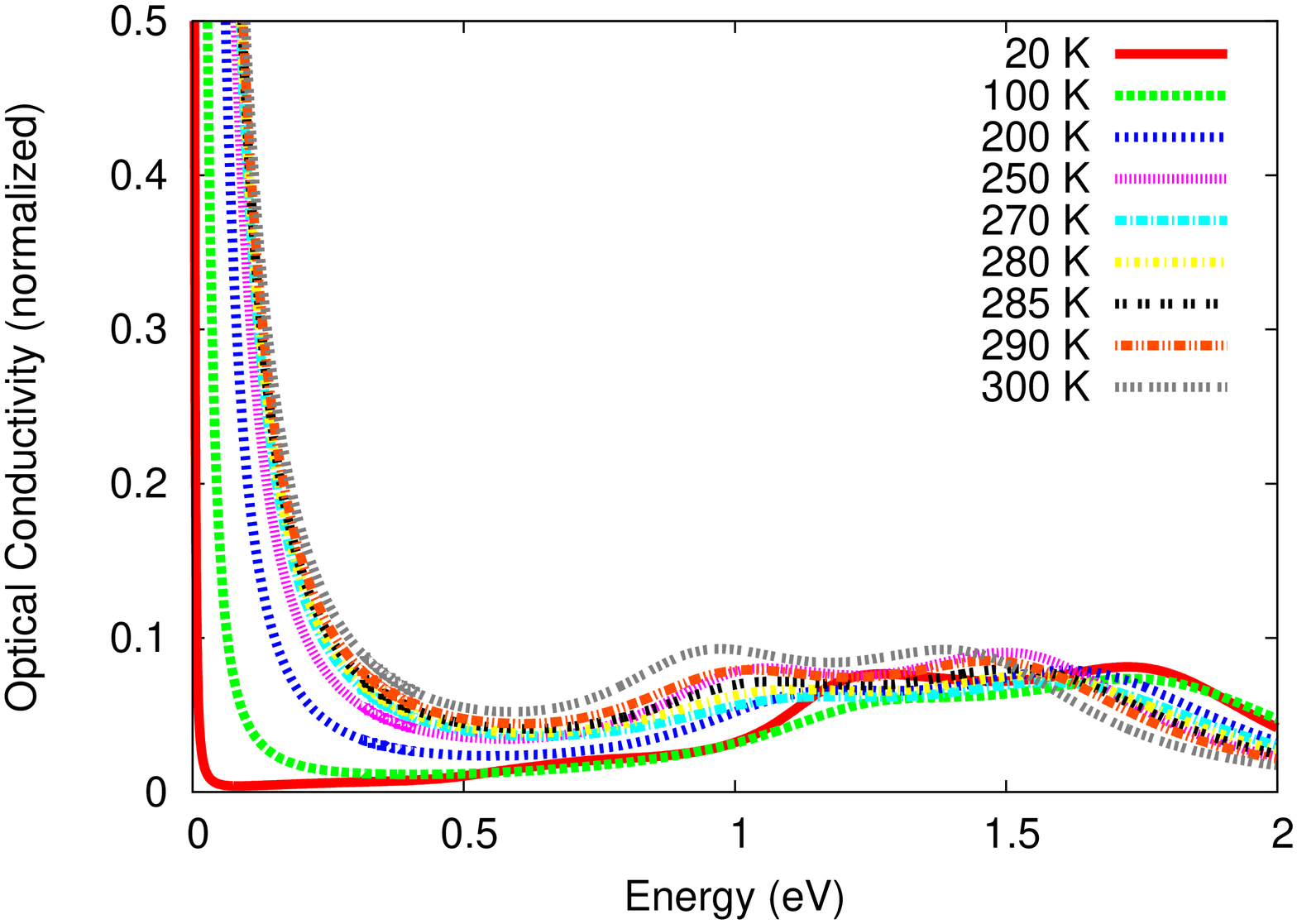}

(b)
\includegraphics[angle=270,width=0.8\columnwidth]{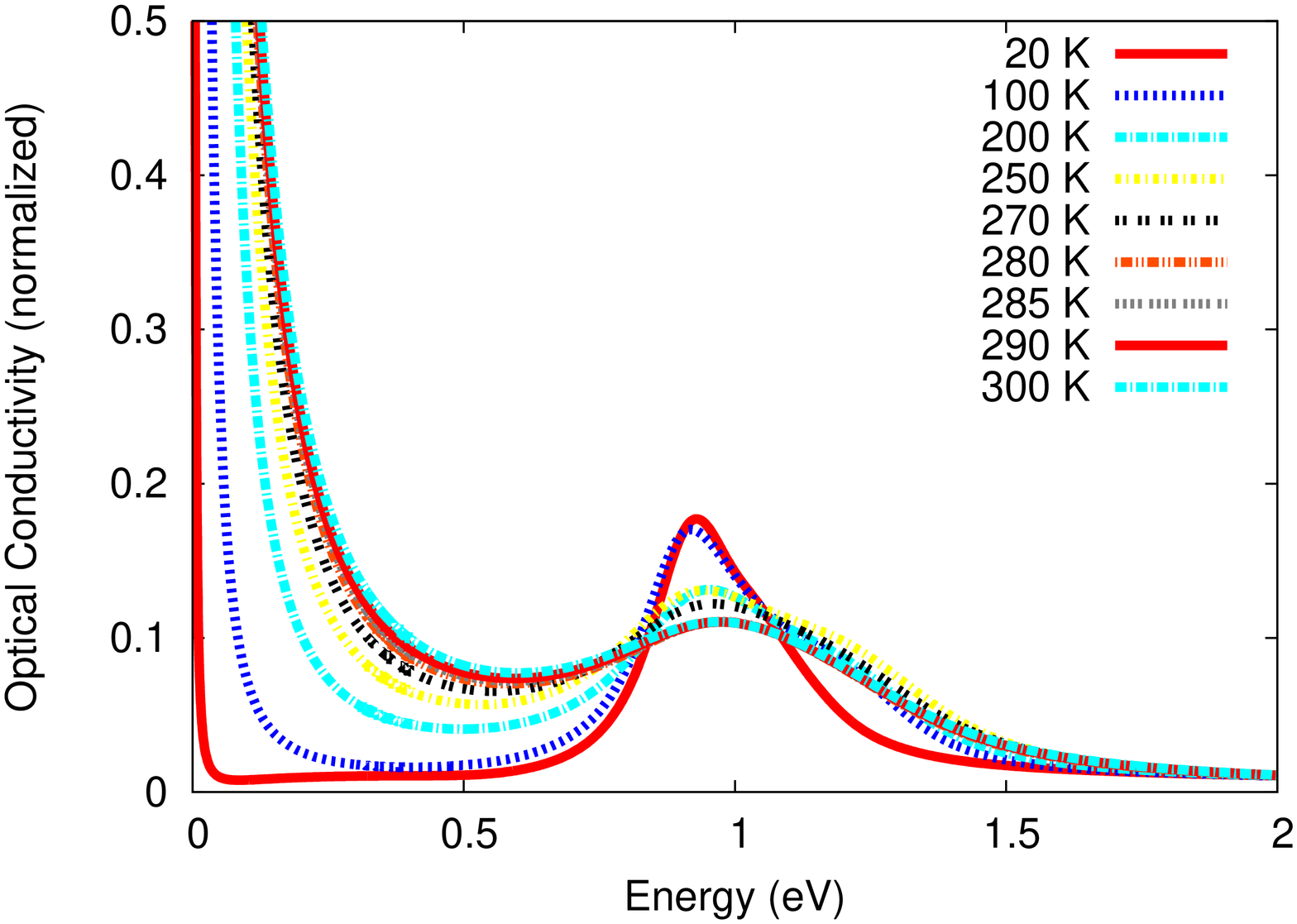}
\caption{(Color Online) DMFT results for $T$-dependent optical conductivity when
 (a) temperature increasing from 0 K to 300 K, (b) temperature decreasing from 
300 K to 0 K.}
\label{fig5}
\end{figure}
\begin{figure}
\centering
\includegraphics[angle=270,width=0.8\columnwidth]{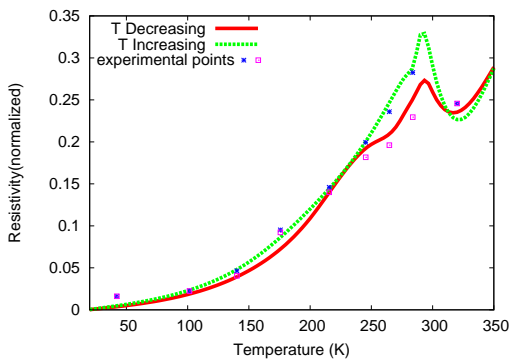}
\caption{(Color Online) Temperature dependent DMFT resistivity for temperature 
increasing and decreasing. DMFT results manifests good agreement with previous 
experimental results (represented by two types of points for temperature 
increasing and decreasing after A. F. Fang et al.\cite{Fang}.}
\label{fig6}
\end{figure}
\noindent I now show how the approach predicted above explains 
a wide variety of physical responses noticeable in IrTe$_2$. 
The DMFT many body density of states for different value of $\alpha$ is shown in 
Fig.\ref{fig2}. I observe that with increasing value of $\alpha$ a pseudogap 
like behavior is emerging in the density of states near Fermi level, 
which in turn proves 
the role of spin orbit coupling in ordered state of IrTe$_2$. Without the 
presence of spin orbit interaction the transition related 
changes is not observed in its theoretical spectra. As displayed in Fig.\ref{fig2} the 
spectral function exhibits a prominent orbital selective pseudogap for small 
value of $\alpha$ 
which confirms the presence of SO interaction tuned ordering in the system 
near 270 K. Since the ordering can be found experimentally also I will take 
$\alpha=0.05$ in further calculation.
Moreover materials with partially filled t$_{2g}$ levels, 
like IrTe$_2$ also, charge degree of freedom is coupled with the orbital 
one, resulting in Peierls instability induced density wave. Presence of these 
type of density wave transition can also be found here. 

\noindent Now the DMFT spectral function for the Ir-d (a band) and Te-p (b band) 
bands is given in Fig.\ref{fig3}a with 
increasing temperature and in Fig.\ref{fig3}b with decreasing temperature and 
corresponding self energies in Figs.\ref{fig4}a, \ref{fig4}b, \ref{fig4}c and 
\ref{fig4}d. 
Both the spectral functions with increasing tempearture exhibits a `peak'  
near Fermi energy at lower temperature and a pseudogap like feature 
at Fermi energy for both the bands. For the `a' band a broad peak comes closer to 
the Fermi level with increasing temperature whereas the peak remains pinned near
 Fermi level 
for the `b' band. However with decreasing temperature a sharp ``polelike" 
feature near Fermi 
energy is spotted in `b' band closer to the Fermi level in almost similar 
fashion as earlier but in the other band the sharp pole is observed exactly at 
 the Fermi level unlike the earlier one. This sharp feature is 
originated due to the finite value of $\alpha$ so it may be argued that 
the `pole' structure along with the low energy pseudogap 
is a result of SO coupling. I further find that the sharp pole in the spectral 
function is a van Hove singularity which disapperars at low temperature.
The remarkable shift of the peak from the Fermi level reduces the kinetic energy
 of electrons which may be the driving force of the transition. However due to 
the presence of pseudogap this dichalcogenide shows bad metallic behavior 
throughout.  

\noindent Interestingly the self energies prove the non Fermi liquid behavior (Im $\Sigma (0)\textgreater 0$)exactly at 
the 250 K temperature above which the anomaly in the resistivity can be found. 
The panels on the left show Ir-5d band self energies while the right panels are 
for Te-5p self energies. 
The imaginary part of self energies are 
shown in Fig.4a and in Fig.4b for both the bands with increasing and 
decreasing temperature. A careful scrutiny of the imaginary part of 
self energy for `a' band with increasing temperature till 300 K shows a 
prominent pole in the energy range of 0.2 eV i.e. very near to Fermi level. 
Now this pole remains on Fermi level at temperature near 270 K, same 
as the temperature 
where resistivity shows a jump. Particularly from this temperature 
onwards Im $\Sigma_a (0)$ shows a nonzero value.
The finding of nFL (non Fermi liquid) behavior in self energy beyond 
250 K persist till high temperature. But throughout all the temperature the 
other band shows no such important changes. It exhibits incoherent Fermi 
liquid ($\propto -\omega^2$). 
Surprisingly with decreasing temperature imaginary part of `a' band self energy
exhibits a different type of behaviour. A peak is observed throughout 
the temperature 
region which is completely below the Fermi level though the peak in 
the self energy varies with temperature. The peak comes closer to the Fermi 
level near 250 K where the resistivity shows an exception. Peak-width also 
varies with temperature. It decreases and ceased to almost a single line 
at lower temperature. Real part of the self energies (with increasing and 
decreasing temperature) also exhibits same type of behavior. While the `b' 
band self energy (both imaginary and real part) remains almost featureless, the
other band shows a good variation with temperature and gives a clue to 
hysterisis in resistivity.
With increasing temperature the peak in self energy is above Fermi energy 
and then 
comes towards Fermi energy and finally it crosses Fermi level while with 
decreasing temperature it is below Fermi level with peak position changing 
with temperature and coming nearest to Fermi level near 250 K.    
The results support that the Rashba spin-orbit coupling associated 
with inter-orbital coulomb 
interaction has a remarkable effect on the temperature dependent self energy 
and density of states. It is observed that near 250 K self energy and spectral 
function are changing their weight on Fermi level to show singularity.
This is exactly that temperature where the resistivity shows a sudden departure. 

\noindent Now to support the prediction of Rashba SO coupling DMFT result should explain 
transport as well. In DMFT, this calculation is simplified: it is an excel-
lent approximation to compute transport co-efficients di-
rectly from the DMFT propagators \cite{biermann}, since
(irreducible) vertex corrections rigorously vanish for one-
band models, and turn out to be surprisingly small even
for the multi-band case. 
In Fig.\ref{fig5} I present the DMFT optical conductivity as a function of T. 
A reflection of pseudogap at low T is found which closes rapidly with increasing T. 
The optical conductivity with increasing temperature is shown in Fig.\ref{fig5}a. 
It shows a featureless broad peak at high energy upto 1.5 eV with a Drude 
like peak at low energy. Both of the Drude peak and broad peak 
decreases with increasing temperature with large spectral weight transfer to 
higher energies. Since I have kept only two bands nearest to the 
 Fermi energy in LCAO
 calculation, agreement at higher energies is not expected. Since the self 
energy shows nFL behavior the peak at zero energy can be explained as 
a reflection of reduced incoherence in the ordered state at low temperature. 
Moreover a broad peaklike structure can also be found in earlier experiments\cite{Fang}. 
Optical conductivity with decreasing temperature is shown in Fig.\ref{fig5}b. In contrast to the earlier one a 
sharp peak is observed in optical conductivity at 1.0 eV which is 
sharper at lower temperature and the peak height also decreases with increase 
in temperature and becomes broader. This change in peak is directly 
coming from the Rashba SO coupling. 
Such a temperature dependence of optical conductivity over a wide energy region 
is quite unconventional and reminiscent of phase transition in the system. 
Clean orbital selective singularity found to be set in with decreasing 
temperature via rapid spectral weight transfer. 
This type of pseudogap is a 
precursor to density wave in this system. The overall change in the optical 
conductivity points out the ordering to set in the system.
Here also the zero energy relatively 
sharp peak is found which also decreases significantly with the temperature and 
reflects the reduced incoherence in 
optical conductivity with lowering temperatures.  
 
\noindent Fig.\ref{fig6} shows the temperature dependent DMFT resistivity of IrTe$_2$. Consistent 
with previous experimental\cite{Fang} and theoretical reports \cite{Yang} pure IrTe$_2$ 
exhibits a hysteresis in resistivity. With increasing temperature resistivity 
shows a peak near about 280 K and with decreasing temperature resitivity 
shows a peak at about 270 K which agrees well also with earlier 
experiments\cite{Fang}. It shows a sudden increase in resistivity upon 
cooling. The resistivity shows bad metallic nature without any anomaly 
upto lowest temperature below this hysteresis region. The bad metallic nature 
can be explained as opening of pseudogap alongside a van Hove 
singularity at Fermi level. Since the resitivity reflects bad metal so the 
consequent short scattering mean free path makes quasiclassical Boltzmann 
equation based transport calculation invalid and in this regime DMFT should fit 
best. This bad metallic behavior below hysterisis region is due to 
reduction of the strong normal state scattering. 
This type of behavior in resistivity is reminiscent of 
structural transition. 
\begin{figure}
\centering
(a)
\includegraphics[angle=270,width=0.4\columnwidth]{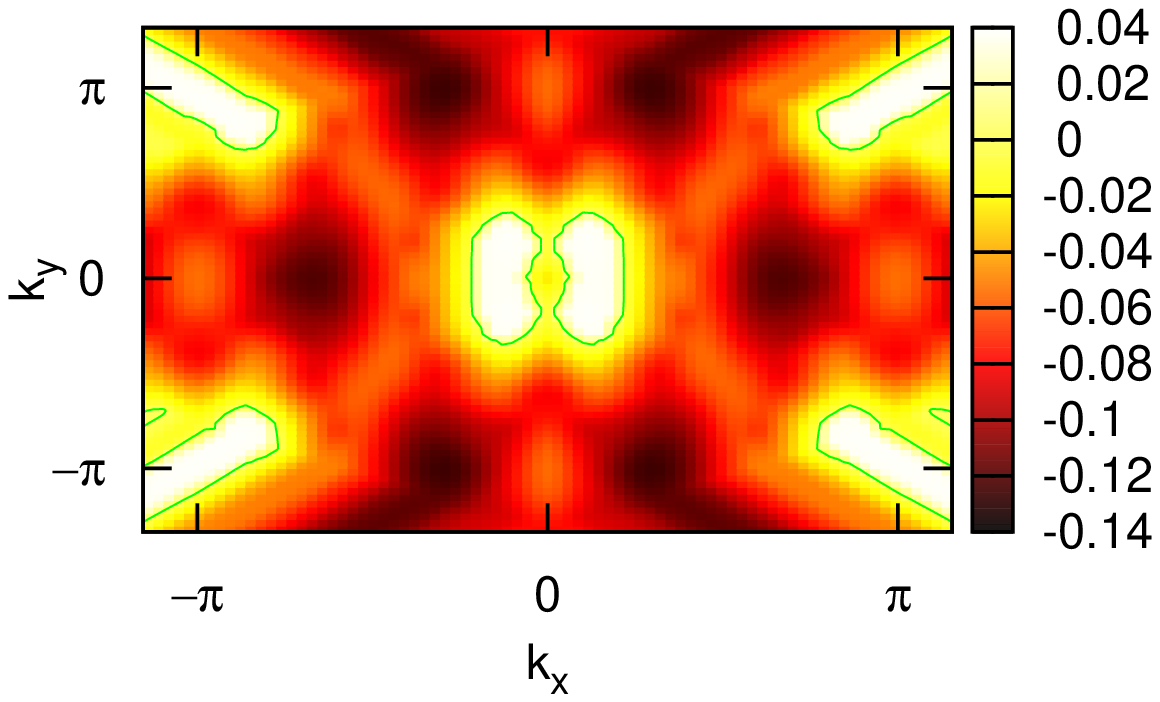}
(b)
\includegraphics[angle=270,width=0.4\columnwidth]{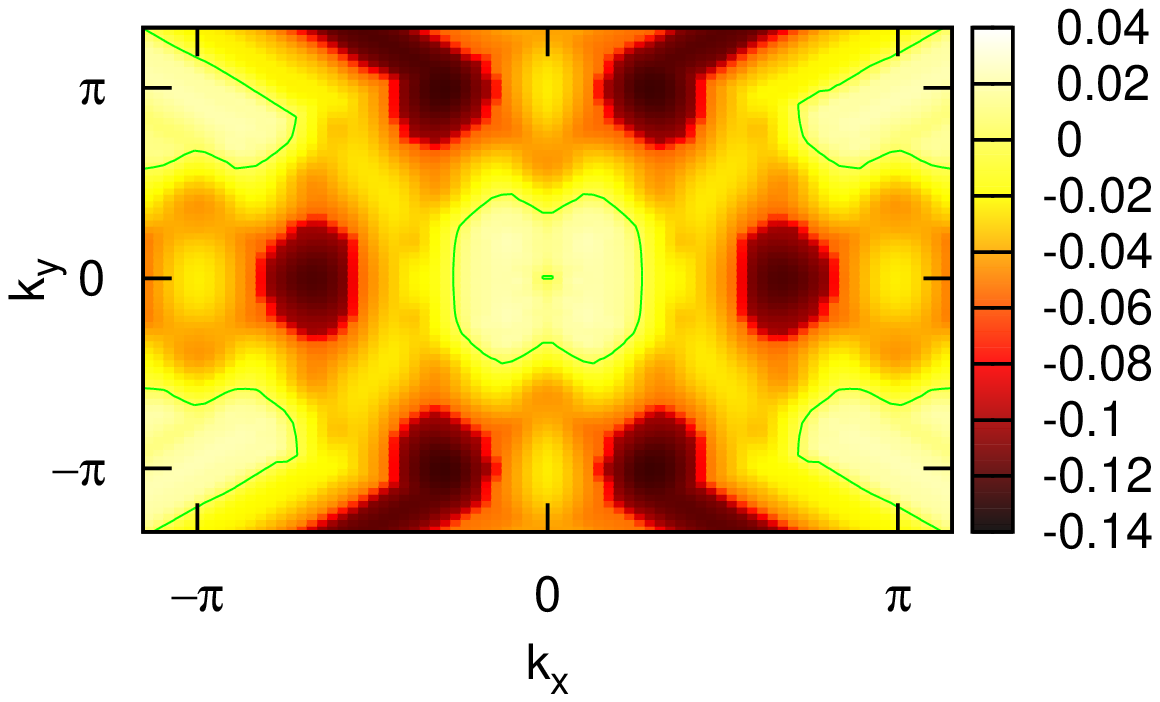}

(c)
\includegraphics[angle=270,width=0.4\columnwidth]{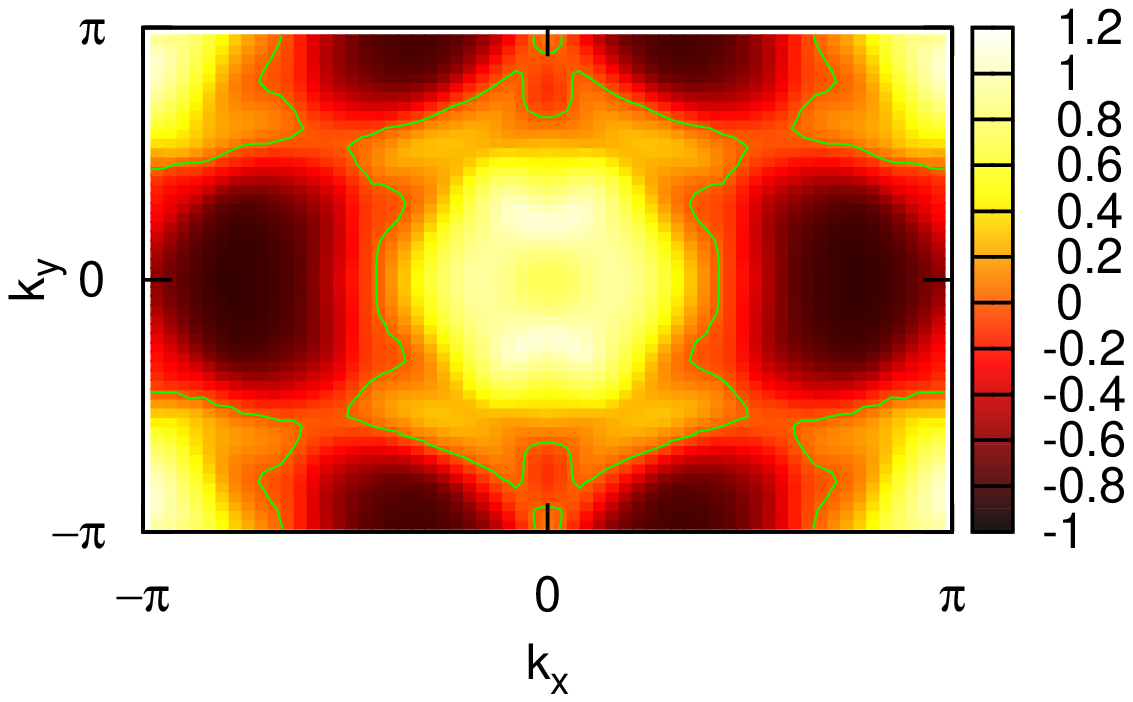}
(d)
\includegraphics[angle=270,width=0.4\columnwidth]{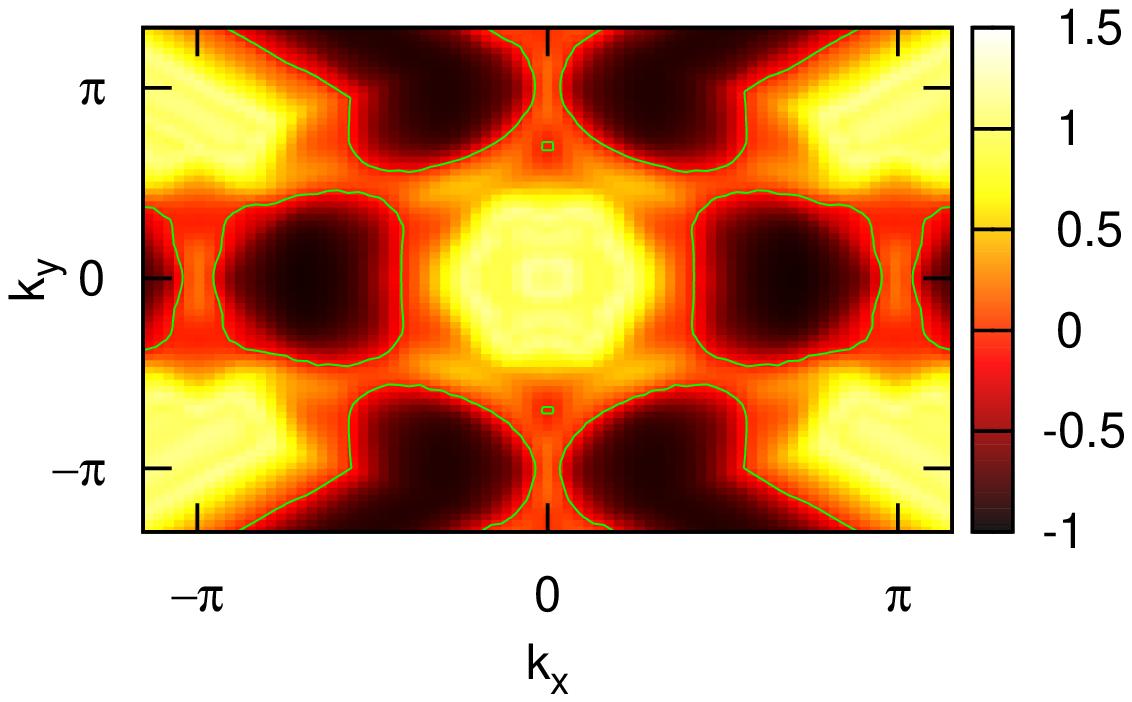}
\caption{(Color Online)  DMFT FS spectra for $\Gamma$-M-K-$\Gamma$ and A-L-H-A direction at temperature
20 K ((a) and (c)) and at 290 K ((b) and (d)) while temperature is increasing.}
\label{fig7}
\end{figure}
\begin{figure}
\centering
(a)
\includegraphics[angle=270,width=0.4\columnwidth]{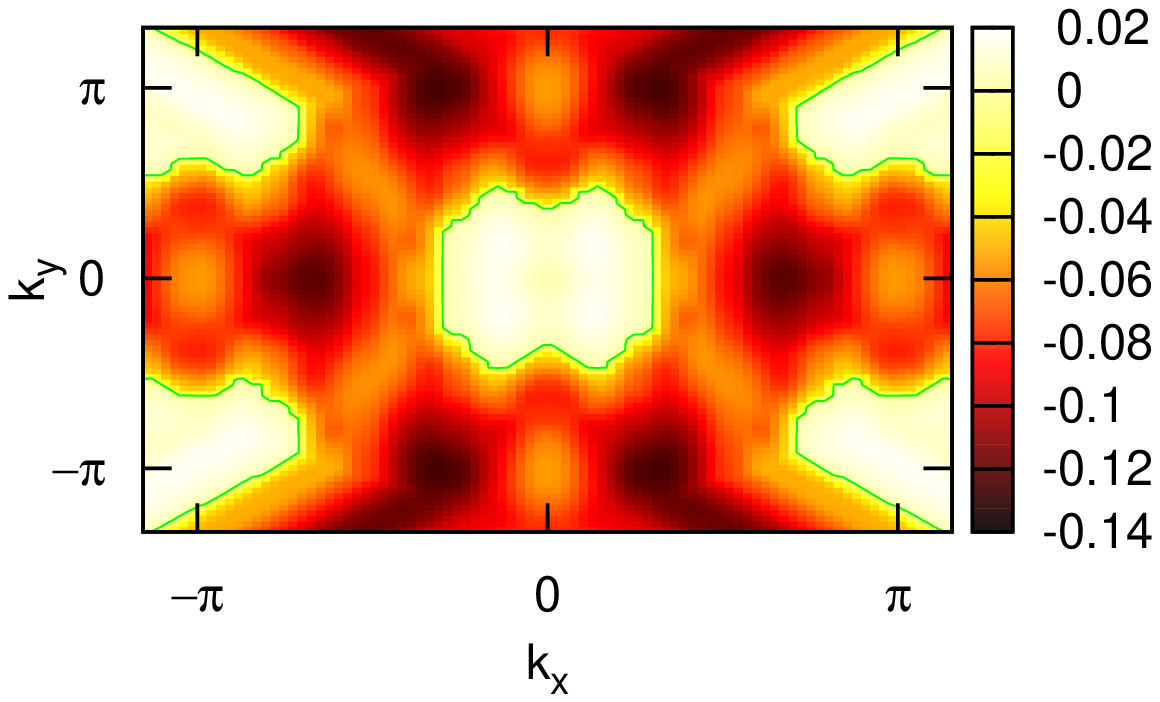}
(b)
\includegraphics[angle=270,width=0.4\columnwidth]{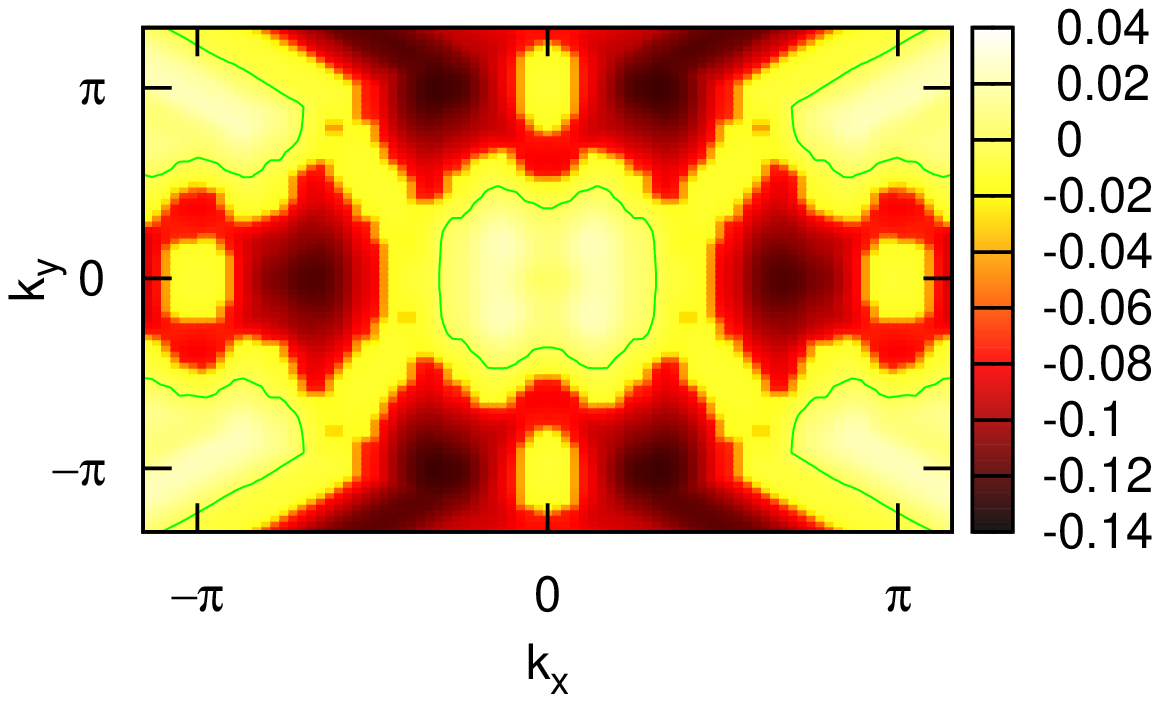}

(c)
\includegraphics[angle=270,width=0.4\columnwidth]{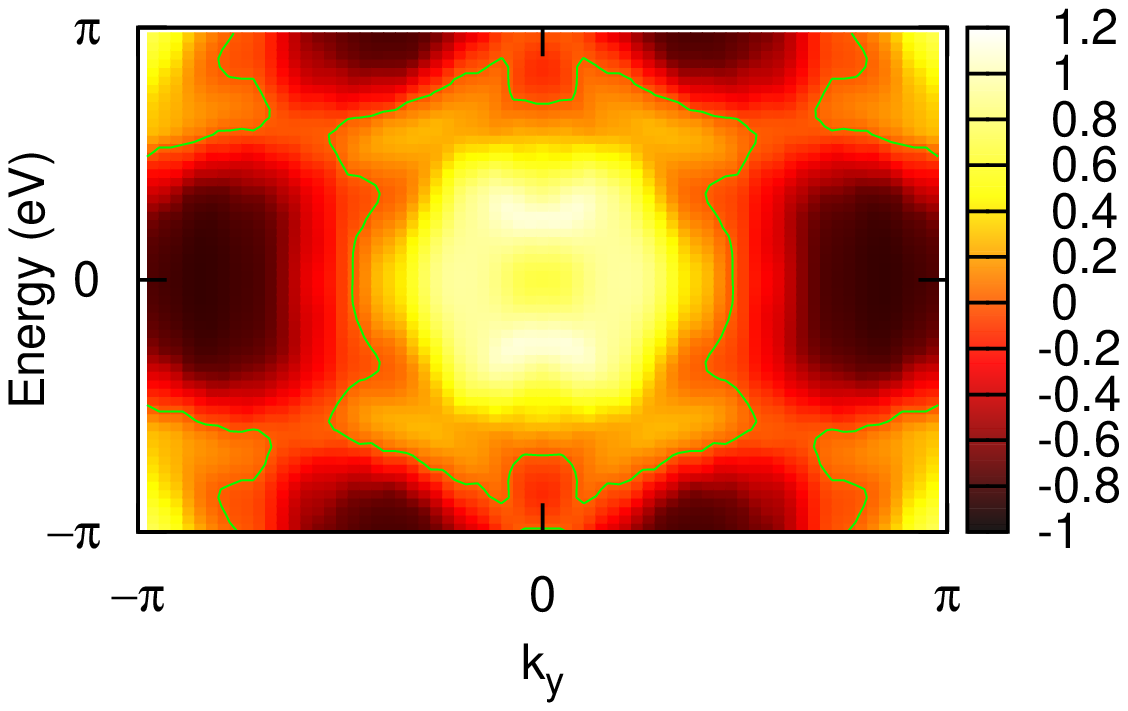}
(d)
\includegraphics[angle=270,width=0.4\columnwidth]{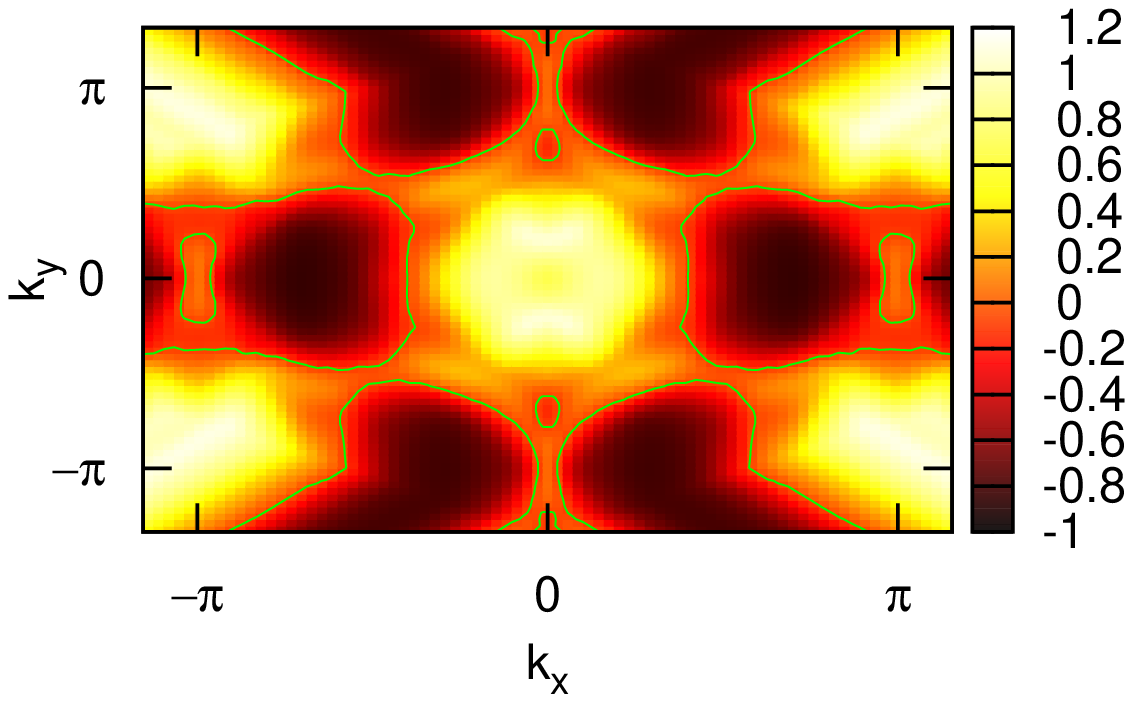}
\caption{(Color Online)  DMFT FS spectra for $\Gamma$-M-K-$\Gamma$ and A-L-H-A 
direction at temperature
20 K ((a) and (c)) and at 290 K ((b) and (d)) while temperature is decreasing.}
\label{fig8}
\end{figure}

\noindent Thus the transport and spectral properties undergoes very good accord with 
experiments which supports the proposal of Rashba spin orbit coupling in 
IrTe$_2$. If the Rashba spin orbit coupling alternative is the cause for this 
phase transtion then the one particle spectral function and renormalized 
band dispersion will also give the proof for same. 
DMFT FS spectra along A-L-H-A direction are also shown (Fig.\ref{fig7} and Fig.
\ref{fig8}) to see 
the change in 
three dimensional FS map for temperature increasing and decreasing. 
Our DMFT FS spectra are consistent with the earlier results \cite{ootsuki}. 
In low temperature near 20 K 
the FS spectra (Fig.\ref{fig7}a and Fig.\ref{fig8}a) shows almost same behavior while 
at higher temperature the FS 
spectra becomes different. At 290 K it shows different map though 
the symmetry remains same.
The FS maps are calculated along the $\Gamma$M-K-$\Gamma$ and A-L-H-A direction 
of the brillouin zone (BZ).
Ir-Ir (Ir-Te) bond in the crystal 
structure corresponds to the A-L (A-H) direction. The centre of the BZ 
corresponds to $\Gamma$ point (kz=0) or A point (kz$\sim$ 0.8 $\pi$). 
At high temperature due
 to thermal excitations relatively large Fermi pockets can be seen (Fig.\ref{fig8}c).
This sizable high-T spreading of FS is due to non zero value of self energy at $\omega =0$.
The flower-shaped Fermi surface is consistent with the earlier calculation\cite{
ootsuki,qian}. Along A-L-H-A direction six connected beads can also be 
identified in the FS map at 290 K which seems to be smeared out at low 
temperature. It is noteworthy that the FS map at 20 K, well below the 
transition temperature (shown in Fig.\ref {fig7}a, b, Fig.\ref {fig8}a and b) is 
appreciably showing almost same structure.  
In contrast the Fermi pockets which are observed at higher temperature are 
vanished at low-T FS map.   
So across the 270 K ordering transition 
the contour of the FS does not change significantly while theFermi
 pockets change dramatically at low temperature. This vanishing of pockets are
 observed perpendicular to the A-H (Ir-Te bond) direction. Therefore it can 
be concluded that both of the orbitals (Ir 5d and Te 5p) are involved 
in this ordering and the transition can be called as a charge-orbital 
ordered state. Further a closer scrutiny of the DMFT results show that the 
central pocket around $\Gamma$ point has developed an elongated shape instead of
 an hexagonal shape as in LDA results. I believe that this important 
modification arises due to an orbital dependent electronic structure reconstruction that 
is developed from the spin orbital coupling factor. 
\begin{figure}
(a)
\includegraphics[angle=270,width=0.4\columnwidth]{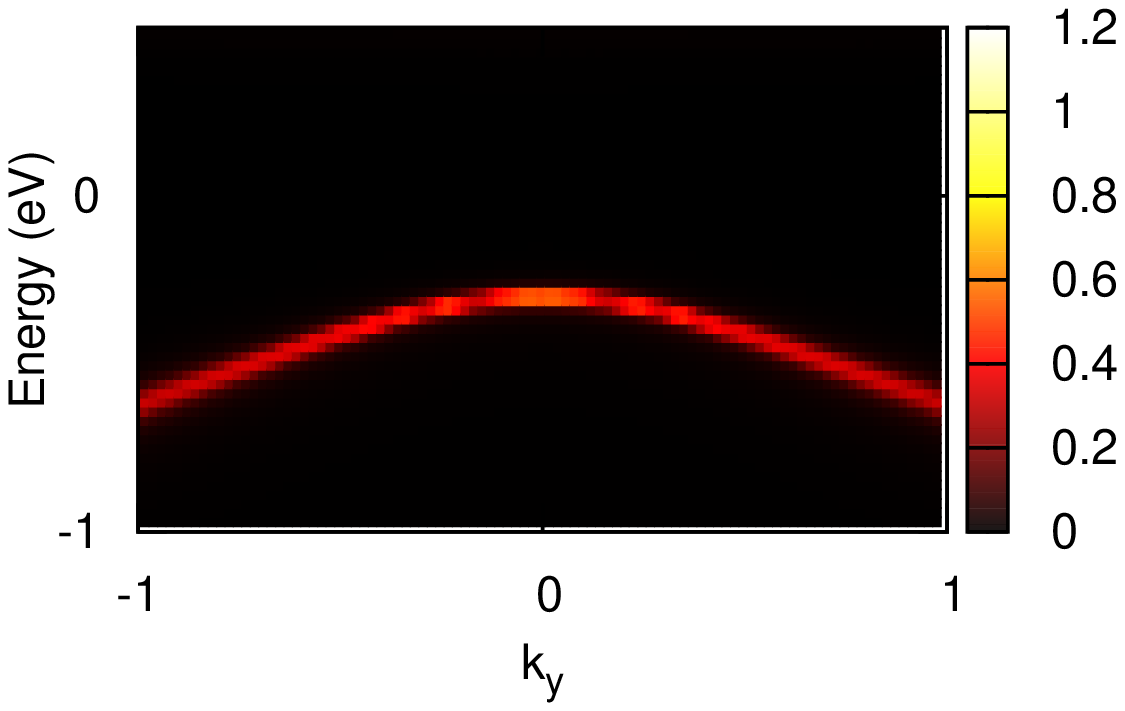}
(b)
\includegraphics[angle=270,width=0.4\columnwidth]{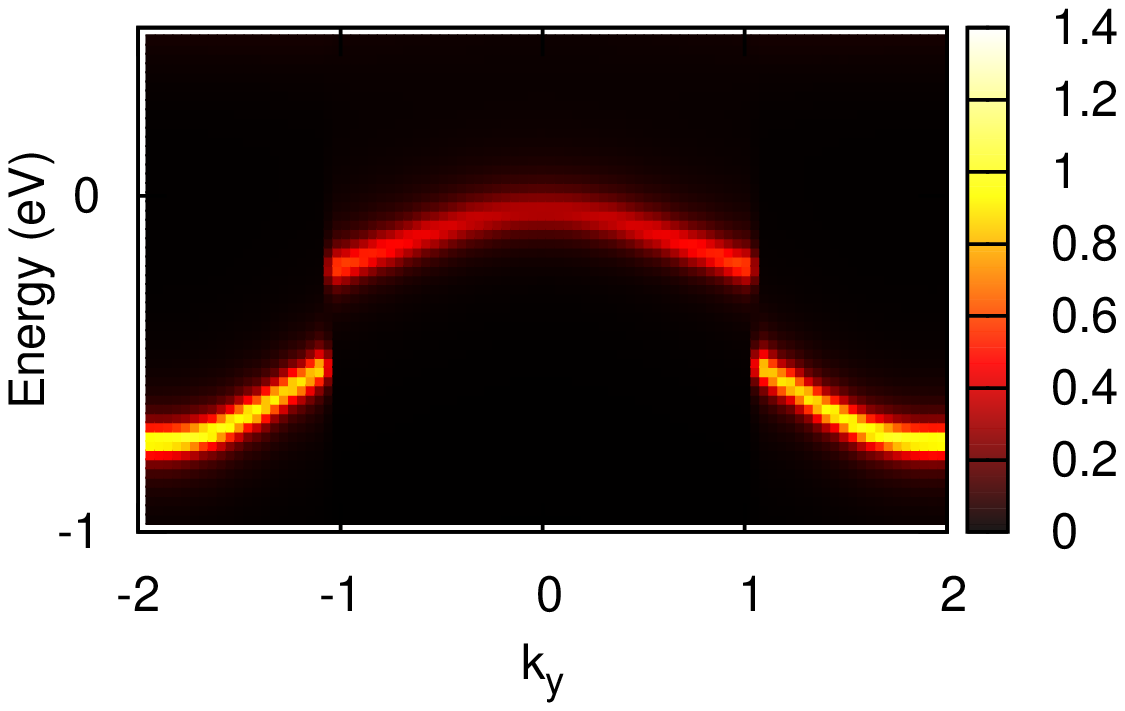}
\caption{(Color Online) Theoretical ARPES map along H-L-H direction at (a) low 
temperature (20 K) and (b) high temperature (290 K). Zero energy along y-axis 
represent Fermi level.}
\label{fig9}
\end{figure}
It can be noted that three dimensional Fermi surface of IrTe$_2$ reveals 
rich dispersions 
reflecting the importance of layered electronic structure.  
This result exactly goes with the previous theoretical and 
experimental band structure results, where, the 
change in FS is proposed as change in 
crystal structure \cite{matsumoto} due to the ordering below 270 K. 
Thus such a qualitative agreement between the DMFT results based on SO coupling 
and earlier theoretical and experimental FS in details gives a credibility of 
the theory. 

\noindent However to give further evidence of SO coupling induced pseudogap 
the same formulation must also describe ARPES as well. I now consider the 
DMFT ARPES maps of band-structure along L-H direction in the brillouin zone at 
both high and low temperature. The dramatic change in the FS is vanishing of 
Fermi pockets at low temperature. To demonstrate this I compare the band 
dispersions along L-H direction at 20 K (Fig.\ref{fig9}a) and 290 K 
(Fig.\ref{fig9}b). It is found that while the band crosses Fermi level at 290 K forming 
the FS pockets, it is well below Fermi level at 20 K leading to a energy 
gap. This gap is not found everywhere in the FS so in the spectral function also
 there is presence of pseudogap. This specifies that the change in FS arises 
from a relocation of the band which also allows a ordering transition. Now this 
related band that forms the pockets is mainly dominated by Te-p orbitals. As 
observed in Fig. \ref{fig3}a, the Te-p DOS exhibits a distinct peak located at 
Fermi level at higher temperature and the peak shifts from Fermi level at lower 
temperature. The energy shift of Te-p band from Fermi level alongwith a 
pseudogap `pinned' to Fermi level of Ir-d band reduces the spectral weight 
at Fermi energy and so the energy of the elctrons also decreases significantly, 
which further drives a phase transition in the system.

\noindent In conclusion, these results show that all the important physical responses 
in IrTe$_2$ and the density wave type transition are associated with the van 
Hove singularity and pseudogap at Fermi level. The bands near Fermi level are strongly 
reallocated forming a pseudogap and on the other hand removing van Hove 
singularity from the Fermi energy. This pseudogap is also reflected in optical 
conductivity which also gives a clear manifestation of interband transition due 
to Rashba spin-orbit coupling. Optical conductivity also shows that carrier 
number reduction due to spin orbit coupling is less than the reduction of 
incoherent scattering. However the steep jump in resistivity at transition 
temperature is also related with nFL behavior due to the pseudogap at the Fermi 
level. Finally the FS and ARPES also supports the pseudogap nature at a 
particular direction of Brillouin zone. This suggests that the charge and 
orbital degree of freedom is coupled to drive the system into a ordering 
transition.
This 
findings also have further important implications on a broader level 
for the systems with partially filled t$_{2g}$ level and 
large SO coupling.    

\noindent SK acknowledges useful discussion and collaboration on
similar systems with Arghya Taraphder.


\begin{thebibliography}{59}
\bibitem{at1}A. Taraphder et al., Phys. Rev. Lett.~{\bf 106}, 236405 (2011).
\bibitem{tokura}Y. Tokura, and N. Nagaosa, Science {\bf 288}, 5465 (2000).
\bibitem{sipos}A. F. Kusmartseva et al., Phys. Rev. Lett.~{\bf 103}, 236401 (2009).
\bibitem{sk}S. Koley,  	arXiv:1606.02841.
\bibitem{shen}Z-X Shen and S. D. Dessau, Physics Reports {\bf 253}, 1 (1995).
\bibitem{motohashi}T. Motohashi et al., Phys. Rev. B {\bf 67}, 6 (2003).
\bibitem{yokoya}T. Yokoya et al., Science {\bf 294}, 5551 (2001).
\bibitem{imada}M. Imada, {\it et al.}, Rev. Mod. Phys, {\bf 70}, 1039 (1998).
\bibitem{kontani}H. Kontani, S. Onari, Phys. Rev. Lett,{\bf 104}, 15 (2010).
\bibitem{Johnston}D.C. Johnston, Advances in Physics, {\bf 59}, 6 (2010).
\bibitem{Li}Y. Li et al., New J. Phys. {\bf 12}, 083008 (2010). 
\bibitem{Mazin}I. I. Mazin et al., Phys. Rev. Lett. {\bf 101}, 057003 (2008).
\bibitem{at3}M. S. Laad et al., J. Phys: Condens. Matter {\bf 24}, 232201 (2012).
\bibitem{vojta}V.H. L{\"o}hneysen et al., Rev. Mod. Phys, {\bf 79}, 3 (2007).
\bibitem{varma}CM Varma, Phys. Rev. B, {\bf 55}, 21 (1997).
\bibitem{coleman}P Coleman Physica B: Condensed Matter, {\bf 259}(1999).
\bibitem{read}N. Read, S Sachdev, Phys. Rev. B,{\bf 42}, 7 (1990).
\bibitem{pwa}P.W. Anderson, {\it Nature Phys.} {\bf 2}, 626 (2006).
\bibitem{Pyon}S Pyon, K Kudo, and M Nohara J. Phys. Soc. Jpn. {\bf 81}, 053701 (2012).
\bibitem{kamitani}M. Kamitani et al., Phys. Rev. B Phys. Rev. B,{\bf 87}, 18 (2013).
\bibitem{Fang}A. F. Fang et al., Scientific Reports {\bf 3}, 1153, (2013).
\bibitem{ootsuki}D. Ootsuki et al., J. Phys. Soc. Jpn. {\bf 82}, 093704 (2013).
\bibitem{zhou}S. Y. Zhou et al., Europhysics Letters {\bf 104}, 27010 (2013).
\bibitem{Yang}J. J. Yang et al.,  Phys. Rev. Lett. {\bf 108}, 116402 (2012).
\bibitem{raub}C. J. Raub, V. B. Compton, T. H. Geballe, B. T. Matthias, 
J. P. Maita, and G. W.Hull, Jr.: J. Phys. Chem. Solids {\bf 26}, 2051 (1965).
\bibitem{matthias}B. T. Matthias, T. H. Geballe, and V. B. Compton: Rev. Mod. Phys. {\bf 35}, 1 (1963).
\bibitem{Larson} P. Larson, I. I. Mazin, and D. A. Papaconstantopoulos, Phys. Rev. B {\bf 67}, 214405 (2003).
\bibitem{Qi} X. L. Qi et al., Phys. Rev. Lett. {\bf 102}, 187001 (2009).
\bibitem{haule}J. Dai et al., Phys. Rev. B {\bf 90}, 235121 (2014). 
\bibitem{lonzarich}S. A. Grigera et al., Science {\bf 294} 329 (2001).
\bibitem{kotliar}A. Georges, {\it et al.}, Rev. Mod. Phys.~{\bf 68}, 13 (1996).
\bibitem{anisimov}V. I. Anisimov, et al., J Phys Cond Matt ~{\bf 9}, 7359 (1997).
\bibitem{at2}S. Koley , N. Mohanta and A. Taraphder,  AIP Conf. Proc. 
{\bf 1461}, 170 (2012).
\bibitem{at4}S. Koley et al., Phys. Rev. B {\bf 90}, 115146 (2014).
\bibitem{at5}S. Koley et al., J Phys Cond Matt {\bf 27}, 185601 (2015).
\bibitem{laad}M. S. Laad et al; Phys. Rev. Lett.~{\bf 91}, 156402 (2003).
\bibitem{qian}T.Qian et al., New Journal of Physics {\bf 16}, 123038 (2014).
\bibitem{biermann}J. Tomczak and S. Biermann, Phys. Rev. B 80, 085117
(2009).
\bibitem{matsumoto}N. Matsumoto et al., J. Low. Temp. Phys. {\bf 117} 1129 (1999). 
\end{thebibliography}
\end{document}